\pdfoutput=1
\documentclass[prd,twocolumn,amsmath,amssymb,superscriptaddress,longbibliography]{revtex4-2}
\usepackage{bm}
\usepackage{graphicx}
\usepackage{booktabs}
\usepackage{microtype}
\usepackage{hyperref}
\usepackage{orcidlink}

\begin{document}

\title{Signature Failure and Minimum Spatial Quanta:\\
Shear-Induced Quantum Boundaries for Gravitational Path Integrals}

\author{Edward J. Shaya\,\protect\orcidlink{0000-0002-3234-8699}}
\affiliation{Department of Astronomy, University of Maryland, College Park, USA}
\email{eshaya2@gmail.com}

\date{}

\begin{abstract}
We investigate whether quantum metric fluctuations can destroy the
differentiable Lorentzian support required for spacetime. We conjecture
that non-Lorentzian metric elements are absent from the physical manifold
and restrict admissible histories to a single continuing, causally
connected four-dimensional geometry. To explore this proposal, we
extrapolate the Gaussian transverse-traceless graviton vacuum to an
additive metric ensemble with physical Fourier cutoff $k_{\max}=1/q$.
A lattice model with local repair and selection of periodically winding
differentiable-state components gives a minimum surviving scale
$q_{\min}\simeq1.9\,\ell_{\rm P}$ over the simulated duration.
We then examine two complementary curvature-induced deformations.
For the Schwarzschild electric Weyl field, the smallest boundary radius
at the center of a black hole, occuring when $q\simeq2.5\,\ell_{\rm P}$, is
$r_{\rm qb}\simeq2.62\,C_W^{1/3}(r_g\ell_{\rm P}^2)^{1/3} \simeq1.91\times10^{-22}\,
C_W^{1/3}\left(\rm M/\rm M_\odot\right)^{1/3}\mathrm{m}$,
where $r_g=GM/c^2$ and $C_W$ parameterizes the finite-$q$ Weyl response.
For homogeneous radiation, a supplementary Gaussian four-metric
ensemble subject to linearized de Donder conditions gives
$\rho_{\rm qb}/\rho_{\rm P}\simeq1.31\times10^{-2}/C_R$
at $q=2.5\,\ell_{\rm P}$, where $C_R$ is the corresponding
Ricci-response coefficient. The local homogeneous criterion is
unchanged under reversal of the expansion rate.
These model-dependent thresholds motivate a possible replacement of
classical singular continuations by finite-curvature quantum boundaries.
\end{abstract}

\maketitle

\section{Introduction}
\label{sec:intro}

The gravitational path integral is usually described as a sum over
geometries, but this leaves open which local configurations count as
spacetime.  A classical metric is nondegenerate and Lorentzian, thereby
defining light cones, causal order, and hyperbolic evolution
\cite{ChoquetBruhat1952,ChoquetBruhatGeroch1969,Geroch1970,HawkingEllis1973,Wald1984}.
At sufficiently short scales the metric itself fluctuates quantum
mechanically, and some realizations need not remain Lorentzian.  If more
than one direction becomes timelike, the principal part is ultrahyperbolic
and ordinary codimension-one data no longer define the usual unconstrained
Cauchy problem \cite{Asgeirsson1937,John1938,CraigWeinstein2009}.  Quantum
field theory on curved spacetime likewise assumes a time-oriented
Lorentzian background \cite{Dimock1980,KayWald1991,Wald1994}.  We ask what
happens to the sum over geometries when local quantum realizations cease to
define admissible spacetime elements.

Previous work has considered Euclidean sectors and classical or quantum
transitions between regions of different signature
\cite{HartleHawking1983,EllisSumerukCouleHellaby1992,DrayManogueTucker1991,
Hayward1992,KossowskiKriele1993}.  Those approaches generally construct or
assume a transition, often across a distinguished hypersurface.  Here no
signature-change surface is imposed: local metric realizations fluctuate out
of the Lorentzian class, and we test whether the surviving subset remains a
connected differentiable spacetime.

The idea that spacetime may possess a fundamental discreteness or a
minimum resolvable scale has a long history. Mead showed that combining
quantum localization with gravitational back reaction suggests an
operational lower limit of order the Planck length
\cite{Mead1964}; later minimum-length analyses reached closely related
conclusions \cite{Garay1995,Hossenfelder2013}. String scattering
provides another route: sufficiently energetic attempts to improve
spatial resolution do not continue to resolve arbitrarily short
distances, and the effective resolution turns around near the string
scale \cite{GrossMende1988}. These arguments do not by themselves
specify a unique microscopic spacetime structure, but they motivate
skepticism that ordinary continuum geometry can be extrapolated
to arbitrarily small distances.

Several quantum-gravity programs implement discreteness more directly.
Causal-set theory postulates a locally finite ordered set as the
microscopic replacement for continuum spacetime \cite{Bombelli1987}.
Loop quantum gravity instead finds discrete spectra for geometric
operators such as area and volume, with spin-network states providing a
discrete quantum geometry \cite{RovelliSmolin1995}. Causal dynamical
triangulations define a nonperturbative sum over geometries through
causal assemblies of simplices and studies the emergence of a
macroscopic semiclassical spacetime \cite{Ambjorn2005}. These
approaches differ substantially in ontology and dynamics.

There is also a direct physical reason for expecting a limiting scale
of order the Planck length, and it follows from the virtual excitations
of the vacuum itself.  An excitation confined to a region of
characteristic size $q$ is a state whose reduced Compton wavelength is
$q$, and therefore carries an effective mass $m(q)\sim\hbar/(cq)$,
equivalently an energy $m c^{2}\sim\hbar c/q$.  As $q$ approaches
$\ell_{\rm P}$ this reaches the Planck mass, so Planck-scale curvature
fluctuations are no longer parametrically rare.  The same mass has the
gravitational radius $r_s(q)=2Gm(q)/c^2\sim2\ell_{\rm P}^2/q$, using
$\ell_{\rm P}^{2}=\hbar G/c^{3}$.  The two lengths cross at
\begin{equation}
 q=\sqrt{2}\,\ell_{\rm P},
\label{eq:comptoncrossover}
\end{equation}
which is the familiar Compton--Schwarzschild estimate of the Planck
scale \cite{Mead1964,Garay1995,Hossenfelder2013}.  For
$q<\sqrt{2}\,\ell_{\rm P}$ one has $r_{s}(q)>q$: the virtual excitation
that any attempt to resolve the scale $q$ must contain would close a
region larger than $q$ itself.

Equation~(\ref{eq:comptoncrossover}) only identifies the scale at which
localization becomes self-undermining.  The independent quantitative bound
below instead follows from the loss of differentiable Lorentzian support
under transverse-traceless vacuum fluctuations.

We adopt the following interpretation. A Lorentzian realization remains
part of the manifold, whereas a non-Lorentzian realization is
represented by a null element $\varnothing$ and is absent from the
physical manifold. An elementary inadmissible region will be called a
\emph{quantum topological hole} (QTH). Physically, a QTH is a four-dimensional
quantum-sized excision from the manifold. An isolated QTH is surrounded
topologically by an $S^3$ and changes the topology, and hence the Betti-number
content, of the manifold. For a specified distribution ${\cal Q}$ of QTHs, the gravitational path
integral assigns an amplitude
\begin{equation}
 \Psi[h_{ij},{\cal Q}]
 =\int_{{\cal Q};\,g|_{\partial{\cal M}}=h}
 {\cal D}g\,
 \exp\!\left(\frac{iS[g]}{\hbar}\right).
\label{eq:psi_qth_sector}
\end{equation}
The formal wavefunctional is then obtained by summing over manifolds with
different numbers and distributions of holes,
\begin{align}
 \Psi[h_{ij}]
 &=\sum_{\cal Q}\Psi[h_{ij},{\cal Q}] \\
 &=\sum_{\cal Q}
 \int_{{\cal Q};\,g|_{\partial{\cal M}}=h}
 {\cal D}g\,
 \exp\!\left(\frac{iS[g]}{\hbar}\right).
\label{eq:psi_sum_qth}
\end{align}
Here $\sum_{\cal Q}$ denotes a sum over manifolds with different QTH
content and distributions.  For practical calculations, we use the more
compact notation
\begin{equation}
 \Psi[h_{ij}]
 =\int {\cal D}(g,\varnothing)\,
 \exp\!\left(\frac{iS[g]}{\hbar}\right),
\label{eq:pathintegral_intro}
\end{equation}
where the explicit bookkeeping over ${\cal Q}$ has been suppressed and
$\varnothing$ denotes locations at which no admissible metric element
exists.  The symbol $\varnothing$ is not an additional field value, and
${\cal D}(g,\varnothing)$ is not a new fundamental measure derived here;
it is shorthand for the enlarged configuration space represented formally
in Eq.~(\ref{eq:psi_sum_qth}).  In this form, the sum over geometries assigns
amplitudes to configurations with different metric and topological content.
The connection to the signature calculation below is direct.  Signature is
configuration-diagonal: for a metric-configuration eigenstate, whether the
local four-metric is Lorentzian is an algebraic property of that same
configuration.  If ${\cal L}$ denotes the region of metric-configuration
space having Lorentzian signature, the Born probability on an equal-time
slice is therefore,
\begin{equation}
 P_{\rm L}
 =\frac{\displaystyle\int_{\cal L}{\cal D}(\delta g)\,
       |\Psi[\delta g]|^2}
      {\displaystyle\int{\cal D}(\delta g)\,
       |\Psi[\delta g]|^2}.
\label{eq:bornsignature}
\end{equation}
No additional measurement map is required to turn the wavefunctional into a
signature distribution.  For the free graviton vacuum used below, the
equal-time wavefunctional is a real positive Gaussian functional of the TT
mode amplitudes.  Consequently, sampling the correlated Gaussian metric
field in the numerical calculation is precisely sampling the Born density
$|\Psi_0[\delta g]|^2$ of the free vacuum, after imposing the stated Fourier
cutoff and TT projection.  The Lorentzian/non-Lorentzian classification is
then applied configuration by configuration.  

\section{Quantum Metric Fluctuations and a Minimum Spatial Quantum}
\label{sec:qmin}

We now ask whether the increasing importance of quantum metric fluctuations
at smaller scales implies a minimum physical size for the spacetime quanta.
For each trial scale $q$, some fraction of metric realizations becomes
inadmissible.  The question is whether there exists a smallest $q$ for which
the admissible quanta remain sufficiently numerous and sufficiently
connected that spacetime can maintain differentiable support everywhere.

\subsection{Admissibility of an elementary metric quantum}

We begin with the four-dimensional metric rather than with a choice of
spacelike slicing. A metric quantum is called \emph{admissible} when its
local $g_{\mu\nu}$ is nondegenerate and has Lorentzian signature
$(-,+,+,+)$, and inadmissible otherwise.  If a quantum realization becomes
non-Lorentzian, that elementary metric is removed from the physical
manifold and the corresponding four-dimensional excision is a QTH, as
defined in the Introduction.  Since spacetime can grow, such a null region need not be permanent:
when a later quantum realization is again Lorentzian and neighboring
support is available, it can return to the admissible manifold on the
natural light-crossing timescale $q/c$.  This admissibility condition is
local, algebraic, and invariant under coordinate transformations. It does not exclude an ordinary black-hole
interior merely because familiar coordinates change character across a
horizon; the metric remains admissible wherever the four-metric remains
non-degenerate and Lorentzian.

In the transverse-traceless gauge used below, \(\delta g_{0\mu}=0\), so the physical fluctuations reside entirely in the spatial block. In a local orthonormal background frame,
$ g_{\mu\nu}=\operatorname{diag}(-1,h_{ij}). $
By Sylvester’s law of inertia, the four-metric is Lorentzian if and only if \(h_{ij}\) is positive definite,
so positive definiteness of $h_{ij}$ provides the computational test of
four-dimensional Lorentzian admissibility used below.

The numerical calculation is based on the ordinary quantized
transverse-traceless gravitational field. We give the construction in some
detail because the Gaussian ensemble, its normalization, its spatial
correlations, and its temporal correlations all follow from the quantum
theory rather than being freely chosen noise.

\subsection{Tensor modes and canonical quantization}

For a spatially flat Friedmann background,
\begin{equation}
 ds^2=-c^2dt^2+
 a^2(t)\left[\delta_{ij}+\chi_{ij}(\mathbf{x},t)\right]dx^i dx^j ,
\label{eq:FRWtensor}
\end{equation}
where $\partial_i\chi_{ij}=0$ and $\chi_{ii}=0$.
The two independent components of $\chi_{ij}$ are the physical tensor
degrees of freedom.  In conformal time $d\eta=c\,dt/a$, the
Einstein--Hilbert action expanded to quadratic order is
\begin{equation}
\begin{aligned}
 S_{\rm TT}
 &=\frac{\overline M_{\rm P}^{\,2}}{8}
 \int d\eta\,d^3x\,a^2\\
 &\quad\times\left[
   \chi'_{ij}\chi'_{ij}
   -\partial_k\chi_{ij}\partial_k\chi_{ij}
 \right].
\end{aligned}
\label{eq:TTaction}
\end{equation}
in units $\hbar=c=1$, with
$\overline M_{\rm P}=(8\pi G)^{-1/2}$.
The quantum treatment of tensor perturbations on an expanding Friedmann background has a long history. \citet{Grishchuk1974} first showed that the time-dependent cosmological background parametrically amplifies gravitational-wave vacuum fluctuations, corresponding quantum mechanically to graviton creation. The canonical theory of cosmological perturbations, including its systematic quantum formulation, was subsequently developed and reviewed in detail by \citet{MFB1992}. We follow this standard framework here, specializing it to the equal-time TT vacuum fluctuations needed for the signature calculation.

We expand the field in polarization tensors, with $k$ denoting the comoving wavenumber,
\begin{equation}
\begin{aligned}
 \chi_{ij}(\mathbf{x},\eta)
 &=\sum_{\lambda=+,\times}
 \int \frac{d^3k}{(2\pi)^3}\,
 \chi_\lambda(\mathbf{k},\eta)\\
 &\quad\times e^{(\lambda)}_{ij}(\hat{\mathbf{k}})
 e^{i\mathbf{k}\cdot\mathbf{x}}.
\end{aligned}
\label{eq:poldecomp}
\end{equation}
with
\begin{equation}
\begin{aligned}
 k_i e^{(\lambda)}_{ij}&=0,
 &e^{(\lambda)}_{ii}&=0,\\
 e^{(\lambda)}_{ij}e^{(\lambda')}_{ij}
 &=2\,\delta_{\lambda\lambda'} .
\end{aligned}
\label{eq:polnorm}
\end{equation}
With this standard normalization, the canonical mode variable is \cite{MFB1992}
\begin{equation}
 v_\lambda=\frac{a\overline M_{\rm P}}{\sqrt{2}}\chi_\lambda ,
\label{eq:canonicalv}
\end{equation}
and the action becomes a sum of harmonic oscillators,
\begin{equation}
\begin{aligned}
 S_{\rm TT}
 &=\frac{1}{2}\sum_\lambda
 \int d\eta\,d^3k\\
 &\quad\times\left[
 |v'_\lambda|^2
 -\left(k^2-\frac{a''}{a}\right)|v_\lambda|^2
 \right].
\end{aligned}
\label{eq:oscaction}
\end{equation}
The Euler--Lagrange equation is
$v_k''+(k^2-a''/a)v_k=0$.
The term \(a''/a\) acts as the cosmological pump responsible for the parametric amplification first emphasized by \citet{Grishchuk1974}. In the short-wavelength regime \(k^2\gg |a''/a|\), however, each mode reduces locally to an ordinary free oscillator.
It is exact for a radiation Friedmann background, for which
$a\propto\eta$ and $a''=0$.

The field operator may, therefore, be written
\begin{align}
 \widehat\chi_{ij}(\mathbf{x},\eta)
 =\sum_\lambda\int \frac{d^3k}{(2\pi)^3}
 e^{(\lambda)}_{ij}(\hat{\mathbf{k}})
 \bigg[
 &\chi_k(\eta)\widehat a_\lambda(\mathbf{k})
 e^{i\mathbf{k}\cdot\mathbf{x}}
 \nonumber\\
 +&\chi_k^*(\eta)\widehat a^\dagger_\lambda(\mathbf{k})
 e^{-i\mathbf{k}\cdot\mathbf{x}}
 \bigg],
\label{eq:fieldoperator}
\end{align}
where
\begin{equation}
\begin{aligned}
 \relax[\widehat a_\lambda(\mathbf{k}),
  \widehat a^\dagger_{\lambda'}(\mathbf{k}')]
 &=(2\pi)^3\delta_{\lambda\lambda'}
 \delta^3(\mathbf{k}-\mathbf{k}'),\\
 \widehat a_\lambda(\mathbf{k})|0\rangle&=0 .
\end{aligned}
\label{eq:commutator}
\end{equation}
For a free mode, the positive-frequency vacuum solution is proportional to
\begin{equation}
 v_k\propto \sqrt{\frac{\hbar}{2k}}\,
 e^{-ik\eta}.
\label{eq:vacmode}
\end{equation}
The phase factor in Eq.~(\ref{eq:vacmode}) also gives the characteristic
physical timescale $\tau_k\sim a/(ck)=1/(c k_{\rm phys})$ of a fluctuation
at comoving wavenumber $k$.
Thus, a fluctuation associated with physical scale $q\sim k_{\rm phys}^{-1}$
varies on the natural timescale $\tau_q\sim q/c$.
The product of the oscillator ground states gives the Gaussian vacuum Born
density sampled in Eq.~(\ref{eq:bornsignature}); Gaussianity is therefore
not an independent stochastic assumption.

\subsection{Vacuum amplitude and the physical cutoff}

Using Eq.~(\ref{eq:vacmode}) together with
Eq.~(\ref{eq:canonicalv}), $\chi_k=\sqrt{2}v_k/(a\overline M_{\rm P})$,
the vacuum mode amplitude is
\begin{equation}
 |\chi_k|^2
 =\frac{2}{a^2\overline M_{\rm P}^{\,2}}
   \frac{\hbar}{2k}
 =\frac{\hbar}{a^2\overline M_{\rm P}^{\,2}k}.
\end{equation}
With the polarization normalization in Eq.~(\ref{eq:polnorm}), the
two-polarization dimensionless tensor spectrum is
\begin{equation}
 P_T(k)
 =\frac{2k^3}{\pi^2}|\chi_k|^2
 =\frac{2\hbar k^2}{\pi^2a^2\overline M_{\rm P}^{\,2}}.
\end{equation}
Introducing the physical wavenumber $k_{\rm phys}=k/a$ gives
\begin{equation}
 P_T(k_{\rm phys})
 =\frac{2\hbar k_{\rm phys}^2}{\pi^2\overline M_{\rm P}^{\,2}} .
\label{eq:PTvac}
\end{equation}
Consequently, all dependence on the arbitrarily chosen value of the FRW scale
factor has disappeared.  From this point onward the calculation concerns
the local vacuum metric fluctuations as a function of physical length
scale.  The FRW background is retained as a convenient framework for the
tensor-mode derivation and for the cosmological applications considered
below.
We define the rms fractional spatial-metric fluctuation by
$\Delta_g^2\equiv\langle\chi_{ij}\chi_{ij}\rangle$.
Equivalently, if
$h^{\rm(sp)}_{ij}=a^2(\delta_{ij}+\chi_{ij})$, then
$\delta h^{\rm(sp)}_{ij}=a^2\chi_{ij}$ and
$\delta h_{\rm(sp)}^{ij}=a^{-2}\chi^{ij}$, so the factors of $a$
cancel from the fractional variance.

We use angular Fourier wavenumbers and define the cutoff length $q$ by
$k_{{\rm phys},\max}=1/q$.
Because $k$ is an angular wavenumber, this definition does not mean that
$q$ is the shortest retained wavelength.  Rather,
\begin{equation}
 \lambda_{\min}=\frac{2\pi}{k_{{\rm phys},\max}}=2\pi q.
\label{eq:lambdamin}
\end{equation}
Thus, $q=k_{\max}^{-1}$ is the inverse-angular-wavenumber cutoff length.  In
the lattice calculation below it is also chosen as the cell spacing, so the
shortest retained wave is sampled by $2\pi$ cells per wavelength.  No factor
of $2\pi$ is to be inserted into Eqs.~(\ref{eq:Dgint})--(\ref{eq:Dgq}),
because those equations are written in terms of the angular cutoff
$k_{\max}=1/q$ itself.

Integrating the vacuum spectrum over physical wavenumber gives
\begin{align}
 \Delta_g^2
 &=\int_0^{k_{{\rm phys},\max}}P_T(k_{\rm phys})\,d\ln k_{\rm phys}
   =\frac{\hbar k_{{\rm phys},\max}^2}{\pi^2\overline M_{\rm P}^{\,2}},
\label{eq:Dgint}\\
 \Delta_g(q)
 &=\sqrt\frac{8}{\pi}\,\frac{\ell_{\rm P}}{q}
   =1.59577\,\frac{\ell_{\rm P}}{q},
\label{eq:Dgq}
\end{align}
where $\overline M_{\rm P}^{\,2}=\hbar c/(8\pi G)$ and
$\ell_{\rm P}^2=\hbar G/c^3$.
The scaling in Eq.~(\ref{eq:Dgq}), $\Delta_g\sim\ell_{\rm P}/q$,
is the Planck-over-scale behavior anticipated in quantum geometrodynamics by
\citet{DeWitt1967} and emphasized by \citet{Wheeler1968}; the derivation
above fixes its coefficient for the present TT vacuum and angular-wavenumber
cutoff.  Since $\Delta_g\propto\alpha^{-1}$, the first question is how rapidly
the fraction of non-Lorentzian metric realizations increases as $\alpha$
decreases.  The second is whether there is a minimum Lorentzian fraction
required for the admissible metric quanta to sustain a connected
differentiable manifold.

Writing $q=\alpha\ell_{\rm P}$ gives
\begin{equation}
 \boxed{\Delta_g(\alpha)
 =\sqrt\frac{8}{\pi}\frac{1}{\alpha}} .
\label{eq:Dgalpha}
\end{equation}
The condition $\Delta_g\sim1$ already shows that signature failure
must become important at a length of order $\ell_{\rm P}$, but it does
not define the minimum spatial quantum.  The latter is determined by
loss of sustainable differentiable connectivity in
Sec.~\ref{sec:numerical}.

As a one-point check, Table~\ref{tab:lorfrac} shows the probability that a
single metric realization remains Lorentzian as the rms tensor amplitude
is increased.  These values use the same traceless TT ensemble as the
numerical calculation, but do not yet impose spatial connectivity.  They can
also be checked independently with random-matrix theory.  The one-point
covariance in Eq.~(\ref{eq:onepointcov}) is that of a rotationally invariant
traceless Gaussian orthogonal ensemble (GOE), whose matrix density is
\begin{equation}
 p(\chi)\propto
 \exp\!\left[-\frac{5}{2\Delta_g^2}\operatorname{Tr}\chi^2\right]
 \delta(\operatorname{Tr}\chi).
\label{eq:tracelessGOE}
\end{equation}
Accordingly, the eigenvalue density is
\begin{equation}
 p(\lambda_1,\lambda_2,\lambda_3)\propto
 e^{-5\sum_i\lambda_i^2/(2\Delta_g^2)}
 \prod_{i<j}|\lambda_i-\lambda_j|\,
 \delta\!\left(\sum_i\lambda_i\right),
\label{eq:tracelessGOEeigs}
\end{equation}
the standard GOE Vandermonde form introduced by \citet{Wigner1958} and
systematized by \citet{Dyson1962}; see \citet{Mehta2004}.  Restricting it
to zero trace permits a useful one-dimensional check.  Write
\begin{equation}
 \lambda_a=\sqrt{\frac{2}{3}}\,r
 \cos\!\left(\theta+\frac{2\pi a}{3}\right),
 \qquad a=0,1,2,
\label{eq:tracelessGOEparam}
\end{equation}
so that $\sum_a\lambda_a=0$, $\sum_a\lambda_a^2=r^2$, and the joint
measure is proportional to
$r^4e^{-5r^2/(2\Delta_g^2)}|\sin3\theta|\,dr\,d\theta$.
Because $g_{00}=-1$ and $g_{0i}=0$ in this one-point test, Lorentzian
signature is equivalent to $I+\chi$ being positive definite, or
$\lambda_{\min}>-1$.  Defining
\begin{equation}
 r_*(\theta)=
 \left[-\sqrt{\frac{2}{3}}\,
 \min_a\cos\!\left(\theta+\frac{2\pi a}{3}\right)\right]^{-1},
\end{equation}
the probability reduces to the single angular integral
\begin{equation}
 P_{\rm L}(\Delta_g)=\frac14\int_0^{2\pi}d\theta\,
 |\sin3\theta|\,
 P\!\left(\frac52,\frac{5r_*^2(\theta)}{2\Delta_g^2}\right),
\label{eq:PLGOE}
\end{equation}
where $P(s,x)$ is the regularized lower incomplete gamma function.  This
semi-analytic result is independent of the Monte Carlo sampling and gives
the final column of Table~\ref{tab:lorfrac}.

\begin{table}[t]
\centering
\caption{Probability that a single fluctuating metric remains Lorentzian
as a function of the metric-quantum scale $\alpha=q/\ell_{\rm P}$ and the
corresponding rms tensor fluctuation amplitude $\Delta_g$.  The Monte Carlo
(MC) values are for the traceless spatial TT ensemble with
$\delta g_{0\mu}=0$; $4\times10^{5}$ realizations were used at each
amplitude.  The GOE column is the independent semi-analytic eigenvalue
integral from Eq.~(\ref{eq:tracelessGOEeigs}).}
\label{tab:lorfrac}
\begin{tabular*}{0.98\columnwidth}{@{\extracolsep{\fill}}cccc}
\toprule
$\alpha$ & $\Delta_g$ & $P_{\rm L}$ MC & $P_{\rm L}$ GOE \\
\midrule
3.99 & 0.4 & 1.0000 & 1.0000 \\
2.66 & 0.6 & 0.9998 & 0.9998 \\
2.00 & 0.8 & 0.9881 & 0.9880 \\
1.60 & 1.0 & 0.9197 & 0.9198 \\
1.45 & 1.1 & 0.8592 & 0.8593 \\
1.33 & 1.2 & 0.7865 & 0.7867 \\
1.14 & 1.4 & 0.6295 & 0.6292 \\
1.00 & 1.6 & 0.4835 & 0.4828 \\
\bottomrule
\end{tabular*}
\end{table}

\begin{figure}[t]
\centering
\includegraphics[width=0.98\columnwidth]{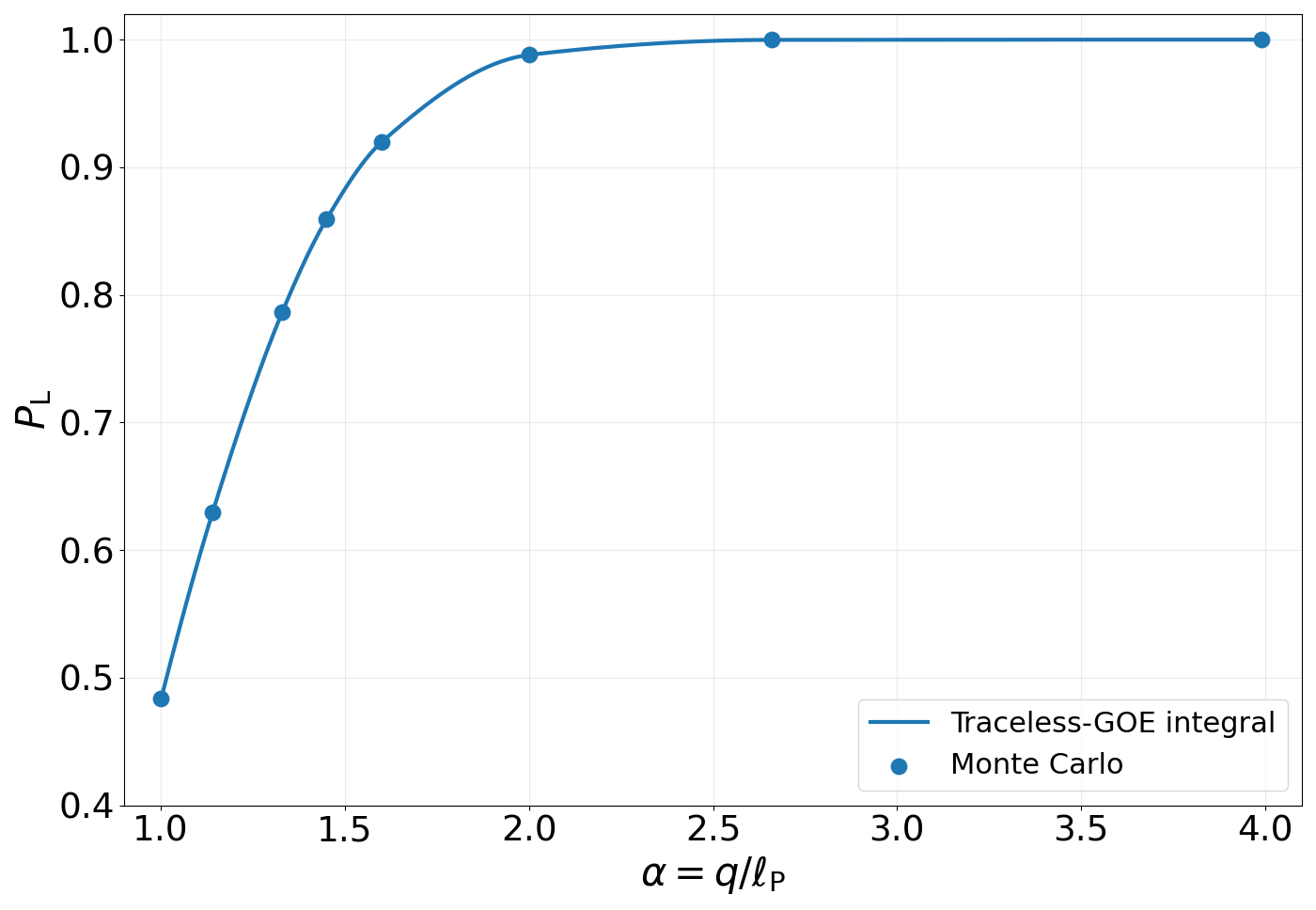}
\caption{One-point Lorentzian probability $P_{\rm L}$ from
Table~\ref{tab:lorfrac}, plotted against the metric-quantum scale
$\alpha=q/\ell_{\rm P}$.  Points show the Monte Carlo calculation and the
solid curve is the independent traceless-GOE eigenvalue integral.  The
corresponding vacuum rms amplitude is
$\Delta_g=\sqrt{8/\pi}/\alpha$.}
\label{fig:lorfrac_plot}
\end{figure}

The same one-point result is plotted in Fig.~\ref{fig:lorfrac_plot}, which
makes the rapid turnover near $\alpha\sim1$--2 more apparent.

The table demonstrates the local mechanism that creates QTHs: as $q$
decreases, Eq.~(\ref{eq:Dgq}) increases the vacuum fluctuation amplitude,
and a growing fraction of individual metric quanta become non-Lorentzian.
The minimum physical $q$, however, is fixed by the correlated field and the
D-state requirement rather than by any one-point probability threshold.

\subsection{Sampling $\chi$ at the TT degrees of freedom}

For each nonzero wave vector, define the transverse projector
$P_{ij}(\hat{\mathbf{k}})=\delta_{ij}-\hat k_i\hat k_j$.
The projector from a symmetric tensor onto its transverse-traceless
part is
\begin{equation}
\begin{aligned}
 {\cal P}_{ij,kl}(\hat{\mathbf{k}})
 =\frac{1}{2}\big(&P_{ik}P_{jl}+P_{il}P_{jk}\\
 &-P_{ij}P_{kl}\big).
\end{aligned}
\label{eq:TTprojector}
\end{equation}
Thus, a trial symmetric Fourier tensor $T_{kl}(\mathbf{k})$ is converted
to the physical tensor field by
\begin{equation}
\begin{aligned}
 T^{\rm TT}_{ij}(\mathbf{k})
 &={\cal P}_{ij,kl}T_{kl}(\mathbf{k})\\
 &=P_{ik}T_{kl}P_{lj}
 -\frac{1}{2}P_{ij}P_{kl}T_{kl}.
\end{aligned}
\label{eq:projectoperation}
\end{equation}
This is the operation implemented in the numerical code.  It removes
the longitudinal and trace components mode by mode, leaving only
the two tensor polarizations.

The equal-time two-point function is
\begin{equation}
\begin{aligned}
 C_{ij,kl}(\mathbf r)
 &\equiv
 \left\langle
 \chi_{ij}(\mathbf x)
 \chi_{kl}(\mathbf x+\mathbf r)
 \right\rangle\\
 &=\int_{k\leq k_{\max}}
 \frac{d^3k}{(2\pi)^3}\,
 {\cal A}(k)\,
 {\cal P}_{ij,kl}(\hat{\mathbf k})
 e^{i\mathbf k\cdot\mathbf r}.
\end{aligned}
\label{eq:spatialcorr}
\end{equation}
where ${\cal A}(k)$ is fixed by
Eq.~(\ref{eq:PTvac}).  Equation~(\ref{eq:spatialcorr}) is why the
lattice cells are not independent random matrices.  They are samples
of one correlated TT field.

After averaging over directions, rotational invariance and tracelessness
give the one-point covariance
\begin{equation}
\begin{aligned}
 \left\langle\chi_{ij}\chi_{kl}\right\rangle
 =\frac{\Delta_g^2}{10}\big(&
 \delta_{ik}\delta_{jl}
 +\delta_{il}\delta_{jk}\\
 &-\frac{2}{3}\delta_{ij}\delta_{kl}\big).
\end{aligned}
\label{eq:onepointcov}
\end{equation}
Consequently,
\begin{equation}
\begin{aligned}
 \left\langle\chi_{11}^2\right\rangle
 &=\frac{2}{15}\Delta_g^2,\\
 \left\langle\chi_{12}^2\right\rangle
 &=\frac{1}{10}\Delta_g^2 .
\end{aligned}
\label{eq:componentvars}
\end{equation}
At $\Delta_g\simeq0.85$, where the local Lorentzian fraction has begun
to decline appreciably, the rms fluctuation of one diagonal component is
only about $0.31$.  

\subsection{Vacuum Field Correlations in Time}

The vacuum field is correlated in time as well as in space.  A free
mode has angular frequency $\omega_k=ck$, so the unequal-time covariance
contains
\begin{equation}
\begin{aligned}
 &\left\langle
 \chi_{ij}(\mathbf x,t)
 \chi_{kl}(\mathbf x',t+\Delta t)
 \right\rangle\\
 &\quad=
 \int_{k\leq k_{\max}}
 \frac{d^3k}{(2\pi)^3}\,
 {\cal A}(k)\,
 {\cal P}_{ij,kl}(\hat{\mathbf k})\\
 &\qquad\times
 e^{i\mathbf k\cdot(\mathbf x-\mathbf x')}
 \cos(ck\Delta t).
\end{aligned}
\label{eq:unequaltime}
\end{equation}
The real symmetrized vacuum correlator is written here because the
numerical calculation samples the Gaussian vacuum distribution as a
stochastic representation of metric configurations.  The important
point is the factor $\cos(ck\Delta t)$: successive configurations are
not independent.

The causal update in Sec.~\ref{sec:numerical} uses $\Delta t=q/c$.  At
the ultraviolet cutoff, $\omega_{\max}\Delta t=ck_{\max}q/c=1$, so even
the shortest retained mode advances by only one radian per
numerical step.  Redrawing an independent TT field after each $q/c$
would, therefore, destroy a real quantum correlation.  The calculation
reported below instead initializes two independent Gaussian quadratures
for every TT Fourier oscillator and evolves them coherently:
\begin{equation}
 \chi_{ij}(\mathbf k,t)
 =A_{ij}(\mathbf k)\cos(ckt)
 +B_{ij}(\mathbf k)\sin(ckt),
\label{eq:quadratureevolve}
\end{equation}
where $A$ and $B$ are independent TT Gaussian fields with the same
vacuum variance.  Equation~(\ref{eq:quadratureevolve}) preserves the
equal-time variance and gives Eq.~(\ref{eq:unequaltime}) for the
discrete mode set.

\section{Numerical Estimate of the Minimum Spatial Quantum}
\label{sec:numerical}

Here we isolate the vacuum problem and ask whether the growing fluctuations
at decreasing $q$, given by Eq.~(\ref{eq:Dgalpha}), eventually destroy
spatial and temporal Lorentzian connectivity.  Curvature-induced changes of
the full four-metric are treated in Sec.~\ref{sec:curvaturestress}.

\subsection{Discrete quantum field}

We generate the perturbations $\chi_{ij}$ of
Eq.~(\ref{eq:quadratureevolve}) by constructing the Fourier coefficients
\begin{equation}
\begin{aligned}
 A_{ij}(\mathbf{k})
 &=F(k)\,{\cal P}_{ij,kl}(\hat{\mathbf{k}})\,
 T^{(A)}_{kl}(\mathbf{k}),\\
 B_{ij}(\mathbf{k})
 &=F(k)\,{\cal P}_{ij,kl}(\hat{\mathbf{k}})\,
 T^{(B)}_{kl}(\mathbf{k}),
\end{aligned}
\label{eq:ABgeneration}
\end{equation}
where $T^{(A)}_{kl}(\mathbf{k})$ and $T^{(B)}_{kl}(\mathbf{k})$ are the
Fourier transforms of two independent Gaussian symmetric-tensor fields,
${\cal P}_{ij,kl}$ is the TT projector, and $F(k)$ is the vacuum-amplitude
weighting given below.

The cubic lattice spacing is $q=\alpha\ell_{\rm P}$, and a periodic box
with $N^3$ cells has side $L=Nq$.  Its angular wave vectors are
$k_i=2\pi n_i/L$, and only modes satisfying $0<|\mathbf k|\leq1/q$ are
retained.  By Eq.~(\ref{eq:lambdamin}), this corresponds to
$\lambda\geq2\pi q$.  The continuum integral is replaced by
\begin{equation}
 \int \frac{d^3k}{(2\pi)^3}
 \longrightarrow
 \frac{1}{V}\sum_{\mathbf k},
 \qquad V=(Nq)^3 .
\label{eq:sumreplace}
\end{equation}

To generate an isotropic Gaussian symmetric tensor, the code first
draws six independent real fields in a Frobenius-orthonormal symmetric
tensor basis.  The three diagonal coefficients have unit variance and
the three off-diagonal matrix entries have variance $1/2$, so that
\begin{equation}
 \|T\|_F^2
 =T_{11}^2+T_{22}^2+T_{33}^2
 +2(T_{12}^2+T_{13}^2+T_{23}^2)
\end{equation}
has isotropic unit covariance in the six-dimensional symmetric-tensor
space.  After a Fourier transform, each mode is projected using
Eq.~(\ref{eq:projectoperation}), evaluated in code as
$T^{\rm TT}=PTP-P\,{\rm Tr}(PT)/2$.

With the orthonormal discrete Fourier convention used in the program,
each projected TT coordinate is multiplied by
$F(k)=\sqrt{16\pi/(kq^3)}$ in Planck units.  Since the TT projector leaves two unit-variance
coordinates, the expected Frobenius variance of one retained lattice
mode is
\begin{equation}
 \left\langle
 |\chi_{ij}(\mathbf k)|_F^2
 \right\rangle
 =\frac{32\pi}{kq^3}.
\label{eq:latticemodevar}
\end{equation}
The expected finite-box rms is therefore
\begin{equation}
 \Delta_{g,{\rm lat}}^2
 =\frac{1}{N^3}
 \sum_{0<k\leq1/q}\frac{32\pi}{kq^3},
\label{eq:latticedelta}
\end{equation}
which approaches the continuum result
$\Delta_g^2=(8/\pi)\ell_{\rm P}^2/q^2$ as the mode sampling becomes
dense.  No post-generation rescaling is applied.

Two independent sets of these projected Fourier coefficients are
generated at the beginning of a trajectory.  They form the vacuum
quadratures $A_{ij}(\mathbf k)$ and $B_{ij}(\mathbf k)$ in
Eq.~(\ref{eq:quadratureevolve}).  At numerical step $n$,
$t_n=nq/c$, each retained mode is
\begin{equation}
 \chi_{ij}(\mathbf k,t_n)
 =
 A_{ij}(\mathbf k)\cos(kqn)
 +B_{ij}(\mathbf k)\sin(kqn).
\label{eq:discretetime}
\end{equation}
An inverse Fourier transform then produces the six real spatial
components at that time.  Because the sine and cosine factors depend
only on $|\mathbf k|$, the Hermitian symmetry of each real quadrature is
preserved.  This construction keeps the discrete vacuum time
correlation rather than replacing it with independent draws.

\subsection{From the quantum field to $P_{\rm raw}$}

The common Friedmann factor does not affect signature, so at every
lattice site we construct the dimensionless spatial matrix
\begin{equation}
 H(\mathbf x,t)
 =I+\chi(\mathbf x,t)
 =
 \begin{pmatrix}
 1+\chi_{11} & \chi_{12} & \chi_{13}\\
 \chi_{12} & 1+\chi_{22} & \chi_{23}\\
 \chi_{13} & \chi_{23} & 1+\chi_{33}
 \end{pmatrix}.
\label{eq:rawmatrix}
\end{equation}
Because $\delta g_{0\mu}=0$ in the TT gauge, the corresponding
four-metric is Lorentzian when $H$ is positive definite.  The
code diagonalizes this real symmetric $3\times3$ matrix and calls the
site raw-Lorentzian when
\begin{equation}
 \lambda_{\min}[H(\mathbf x,t)]>0 .
\label{eq:rawcriterion}
\end{equation}
If $N_{\rm raw}(t)$ sites satisfy
Eq.~(\ref{eq:rawcriterion}), then
\begin{equation}
 \boxed{
 P_{\rm raw}(t)=\frac{N_{\rm raw}(t)}{N^3}
 } .
\label{eq:Praw}
\end{equation}
The correlated quantum-field realizations of $P_{\rm raw}$ are averaged over
an ensemble of trajectories.  Table~\ref{tab:lorfrac} is the corresponding
one-point check.

\subsection{Differentiable support and causal persistence}

Positive definiteness at one point is insufficient for a local
classical geometry because curvature requires second derivatives.  A
centered finite-difference derivative, for example,
\begin{equation}
 \frac{\partial^2H}{\partial x^2}\bigg|_i
 \simeq
 \frac{H_{i+1}-2H_i+H_{i-1}}{q^2},
\label{eq:seconddifference}
\end{equation}
requires support on both sides, while mixed derivatives require the
diagonal neighbors.  We therefore define a D-state site to be an
active site at the center of a complete active
$3\times3\times3$ neighborhood.  A complete local differentiable seed
contains $27$ spatial quanta.

The active/null state is evolved once per physical light-crossing time
$q/c$.  The update is deliberately non-iterative within one time
slice:
\begin{enumerate}
 \item a previously active site survives provisionally only if it is
       raw-Lorentzian on the new slice;
 \item a previously null site can repair only if it is raw-Lorentzian
       and face-adjacent to a site active on the preceding slice;
 \item D-state is evaluated once from this provisional active set;
 \item connectivity is propagated through periodic, face-sharing D-state
       neighbors, and only components with nonzero periodic winding are
       retained;
 \item raw-Lorentzian face-neighbors of the winding D-state set are
       retained as the growth/repair fringe;
 \item all other sites are null until a later causal repair becomes
       possible.
\end{enumerate}
The one-cell repair rule limits propagation of support to at most $q$
during one interval $q/c$.  Newly lost support is not recursively
allowed to erase arbitrarily distant cells during the same interval.
We restrict the sum to histories supporting a single continuing differentiable Lorentzian spacetime. In the numerical model, this requirement is implemented by retaining only spatially winding D-state components on each slice, with local repair from surviving support.

For each $\alpha$, we record $P_{\rm raw}$, $P_{\rm active}$, $P_D$, and
$P_{D,\rm wind}$, where the last quantity is the fraction of the box 
containing D-state components with nonzero periodic winding.
Operationally, a face-sharing path must cross a box face and
return to the same periodic site without crossing a compensating face; in the
unwrapped lattice it therefore returns to a distinct image displaced by a
nonzero integer multiple of the box length. 
The connectivity calculations were performed in periodic boxes with $N=64$
and $N=128$.  Typically three independent
vacuum-field trajectories were evolved for each trial value of $\alpha$,
and each trajectory was followed for 500 time steps, with one step
corresponding to the light-crossing time $q/c$.  A trajectory was classified as forming a quantum
boundary when no periodically winding D-state component remained after 500
steps.  We checked the transition at both box sizes.

We therefore take the present numerical result to indicate
\begin{equation}
 q_{\min}\simeq1.9\,\ell_{\rm P},
 \qquad
 \Delta_{g,c}\simeq0.85 ,
\label{eq:qminprelim}
\end{equation}
As a check on the restriction to TT fluctuations, we repeated the
connectivity calculation with a full four-metric ensemble in which the four
linearized gauge degrees of freedom were projected from the ten independent
components of $g_{\mu\nu}$.  The spanning manifold was lost at a physical
scale of approximately $2.0\,\ell_{\rm P}$, close to the TT result
$q_{\min}\simeq1.9\,\ell_{\rm P}$.  This agreement indicates that the
inferred restriction does not strongly control the minimum spatial scale
to the TT sector.

\section{Curvature-Induced Metric Stress and Quantum Boundaries}
\label{sec:curvaturestress}

The vacuum calculation above asks whether quantum fluctuations alone can
destroy Lorentzian connectivity.  Classical or semiclassical curvature adds
a finite-scale metric stress across an element of size $q$.  We represent
that response schematically by
\begin{equation}
 g^{(q)}_{\mu\nu}
 =
 \eta_{\mu\nu}
 +\chi^{Q}_{\mu\nu}
 +s^{(\mathcal R)}_{\mu\nu},
\end{equation}
where $\chi^{Q}_{\mu\nu}$ is the quantum fluctuation and
$s^{(\mathcal R)}_{\mu\nu}$ is the deformation induced by curvature
across the element.  The general curvature content is displayed by the
irreducible decomposition of the Riemann tensor:
\begin{equation}
\begin{aligned}
R_{\alpha\beta\gamma\delta}
={}&C_{\alpha\beta\gamma\delta}\\
&+\frac12\left(
g_{\alpha\gamma}S_{\beta\delta}
-g_{\alpha\delta}S_{\beta\gamma}
-g_{\beta\gamma}S_{\alpha\delta}
+g_{\beta\delta}S_{\alpha\gamma}\right)\\
&+\frac{R}{12}\left(
g_{\alpha\gamma}g_{\beta\delta}
-g_{\alpha\delta}g_{\beta\gamma}\right),
\end{aligned}
\label{eq:RiemannDecomp}
\end{equation}
where
\begin{equation}
 S_{\mu\nu}\equiv
 R_{\mu\nu}-\frac14 g_{\mu\nu}R .
\label{eq:tracefreeRicci}
\end{equation}
The three contributions are the Weyl tensor
$C_{\alpha\beta\gamma\delta}$, the trace-free Ricci tensor
$S_{\mu\nu}$, and the scalar curvature $R$.  The two applications below
isolate complementary limits.  Schwarzschild spacetime is Ricci flat and
therefore supplies a pure Weyl tidal stress, whereas spatially flat
radiation-dominated Friedmann spacetime has vanishing Weyl tensor and
$R=0$, leaving a pure trace-free Ricci stress.

\subsection{Weyl- and Shear-Induced Quantum Boundaries}
\label{sec:stress}

Anisotropic curvature changes the relative geometry of neighboring metric
quanta and can drive additional cells into the non-Lorentzian sector.  In a
Ricci-flat vacuum spacetime this tidal stress is carried entirely by the Weyl
tensor.  Its kinematic manifestation along a timelike congruence is shear,
which supplies the bridge from curvature to the finite-scale metric
deformation used below.  If sufficiently strong, the resulting proliferation
of QTHs can disconnect the differentiable Lorentzian continuum and form a quantum boundary.

We write the dimensionless conformal spatial metric of an individual
spatial quantum as
\begin{equation}
 H_{ij}(q,t)
 =
 \frac{\gamma_{ij}(q,t)}{a^2}
 =
 \delta_{ij}
 +s_{ij}(q,t)
 +\chi^{Q}_{ij}(q,t),
\label{eq:shearmetric}
\end{equation}
where $s_{ij}(q,t)$ denotes the classical or semiclassical anisotropic
deformation of a metric element of size $q$, and $\chi^{Q}_{ij}(q,t)$ is
the quantum TT fluctuation discussed in Sec.~\ref{sec:qmin}.  A common
isotropic scale factor has been omitted because it cannot change the
signs of the spatial eigenvalues.  By contrast, the anisotropic term
$s_{ij}$ changes the shape of the local metric and can cause its smallest
eigenvalue to pass through zero.

The geometrical shear may be described covariantly using the
trace-free part of the expansion tensor.  For a timelike congruence
with four-velocity $u^\mu$,
\begin{equation}
\begin{aligned}
 \nabla_\nu u_\mu
 ={}&\frac{1}{3}\theta\,h_{\mu\nu}
 +\sigma_{\mu\nu}\\
 &+\omega_{\mu\nu}-u_\mu a_\nu .
\end{aligned}
\label{eq:kinematicdecomp}
\end{equation}
where $\theta$ is the isotropic expansion, $\sigma_{\mu\nu}$ the shear,
$\omega_{\mu\nu}$ the vorticity, and $a^\mu$ the four-acceleration.
The distinction between $\theta$ and $\sigma_{\mu\nu}$ is the same one
that matters in Eq.~(\ref{eq:shearmetric}): isotropic expansion or
contraction rescales the local spatial geometry, whereas shear changes
its shape.  Since the present calculation always concerns one spatial
quantum of size $q$, the natural time over which the local deformation is
evaluated is its light-crossing time $q/c$.  We therefore relate the
anisotropic metric deformation to the local spatial shear by
\begin{equation}
 s_{ij}(q,t)\sim \frac{q}{c}\,\sigma_{ij}(q,t),
\label{eq:sheartos}
\end{equation}
where $\sigma_{ij}$ denotes the spatial components of the shear in the
local orthonormal frame.  This relation supplies the scale-dependent
connection between the shear of the congruence and the deformation
$s_{ij}$ entering Eq.~(\ref{eq:shearmetric}).  Relative to the congruence,
the electric Weyl tensor is
$E_{\mu\nu}\equiv C_{\mu\alpha\nu\beta}u^\alpha u^\beta/c^2$; it is spatial,
symmetric, and trace-free.  We define its dimensionless finite-$q$
deformation by
\begin{equation}
 \mathcal{T}_{ij}^{(W)}(q)\equiv-q^2E_{\hat i\hat j},
\label{eq:tidalstrength}
\end{equation}
where the sign follows the geodesic-deviation acceleration
$D^2\xi^i/D\tau^2=-c^2E^i{}_j\xi^j$.  For a fixed eigenvalue pattern we
write $\mathcal{T}_{ij}^{(W)}=\mathcal{T}Q_{ij}$; the signature-failure
threshold in its scalar amplitude for $q=\alpha\ell_P$ is denoted
$\mathcal{T}_{\rm qb}(\alpha)$.

\subsubsection{Schwarzschild black-hole interiors}
\label{sec:schwarzschild_tidal}

A particularly simple application is the interior of a nonrotating
Schwarzschild black hole.  No matter stress is required: the vacuum
tidal field itself produces radial stretching and transverse
compression of neighboring geodesics.  In a local orthonormal frame, the
relative tidal acceleration across a separation $q$ is of order
\begin{equation}
 \frac{\Delta a_i}{q}
 \sim
 c^2\frac{r_g}{r^3}
 \mathrm{diag}(2,-1,-1),
 \qquad r_g\equiv \frac{GM}{c^2}.
\label{eq:schwarzschildtidal}
\end{equation}
Thus
$E_{\hat i\hat j}^{\rm Schw}=-(r_g/r^3)\operatorname{diag}(2,-1,-1)$.
Over one quantum light-crossing time $q/c$, this produces a shear of
order
\begin{equation}
 \sigma_{ij}^{\rm Schw}
 \sim
 c\frac{r_g q}{r^3}
 \mathrm{diag}(2,-1,-1).
\label{eq:schwarzschildshear}
\end{equation}
Using Eq.~(\ref{eq:sheartos}) then gives the corresponding dimensionless
metric deformation,
\begin{equation}
 s_{ij}^{\rm Schw}(r,q)
 =
 C_W\frac{r_g q^2}{r^3}
 \operatorname{diag}(2,-1,-1).
\label{eq:schwarzschilds}
\end{equation}
Here \(C_W\) parameterizes the finite-\(q\) conversion from electric Weyl curvature to metric deformation; the numerical scans adopt \(C_W=1\).
For Schwarzschild, Eq.~(\ref{eq:tidalstrength}) gives
$\mathcal{T}_{ij}^{(W)}=\mathcal{T}\operatorname{diag}(2,-1,-1)$, where
$\mathcal{T}(r,q)=r_gq^2/r^3$, so that
$s_{ij}^{\rm Schw}\sim\mathcal{T}_{ij}^{(W)}$ in this local model.
The local quantum
metric is therefore
\begin{equation}
 H_{ij}(r,q,t)
 =
 \begin{pmatrix}
 1+2C_W\mathcal{T} & 0 & 0\\
 0 & 1-C_W\mathcal{T} & 0\\
 0 & 0 & 1-C_W\mathcal{T}
 \end{pmatrix}
 +\chi^Q_{ij}(q,t).
\label{eq:schwarzschildH}
\end{equation}
This is a local finite-scale representation of the tidal deformation of
one spatial quantum; it is not an accumulated infall history.
The quantum rms remains
\begin{equation}
\begin{aligned}
 \Delta_g(q)&=\sqrt{\frac{8}{\pi}}\frac{\ell_P}{q}
             =\frac{1.596}{\alpha},\\
 q&=\alpha\ell_P .
\end{aligned}
\label{eq:schwarzschildDg}
\end{equation}
so decreasing radius does not amplify $\Delta_g$.  Instead, increasing
$\mathcal{T}\propto r^{-3}$ shifts the background eigenvalues and increases the
fraction of quantum realizations for which one or more spatial eigenvalues
are nonpositive.  The relevant boundary is reached when those failures are
sufficient to destroy connected D-state support.

We tested this directly with the same correlated TT field and causal
D-state update used in Sec.~\ref{sec:numerical}.  For each $\alpha$ the
Schwarzschild tidal-field strength was increased until connected D-state support
was lost.  The resulting $\mathcal{T}_{\rm qb}(\alpha)$ values are listed in
Table~\ref{tab:schwarzschild_Tqb}:
\begin{table}[t]
\caption{Critical Schwarzschild tidal-field strength $\mathcal{T}_{\rm qb}$ for loss of connected
D-state support at fixed spatial quantum size $q=\alpha\ell_{\rm P}$.
The final column gives
$C\equiv[\alpha^2/\mathcal{T}_{\rm qb}]^{1/3}$, so that
$r_{\rm qb}=C(r_g\ell_{\rm P}^2)^{1/3}$.}
\label{tab:schwarzschild_Tqb}
\centering
\begin{tabular*}{\columnwidth}{@{\extracolsep{\fill}}ccc@{}}
\hline\hline
$\alpha$ & $\mathcal{T}_{\rm qb}$ & $C$ \\
\hline
1.9 & 0.06 & 3.84 \\
2.0 & 0.15 & 3.02 \\
2.2 & 0.25 & 2.70 \\
2.4 & 0.32 & 2.63 \\
2.5 & 0.35 & 2.62 \\
2.6 & 0.38 & 2.62 \\
2.7 & 0.40 & 2.64 \\
3.0 & 0.46 & 2.69 \\
3.5 & 0.54 & 2.83 \\
4.0 & 0.60 & 2.98 \\
5.0 & 0.68 & 3.32 \\
10.0 & 0.84 & 4.91 \\
\hline\hline
\end{tabular*}
\end{table}

\begin{figure*}[t]
\centering
\begin{minipage}{0.48\textwidth}
\centering
\includegraphics[width=\linewidth]{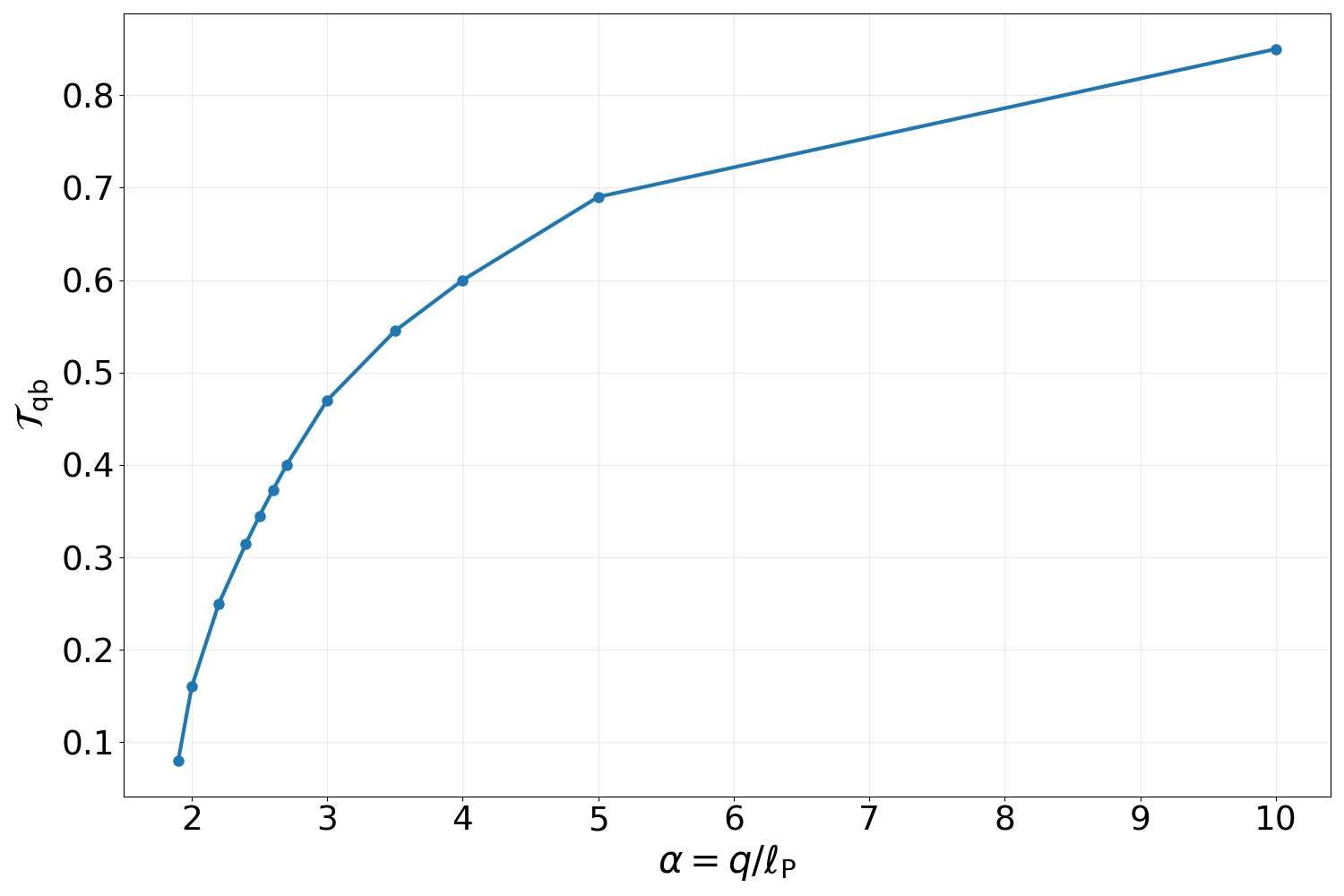}\\[-1mm]
{\small (a) Critical tidal strength}
\end{minipage}\hfill
\begin{minipage}{0.48\textwidth}
\centering
\includegraphics[width=\linewidth]{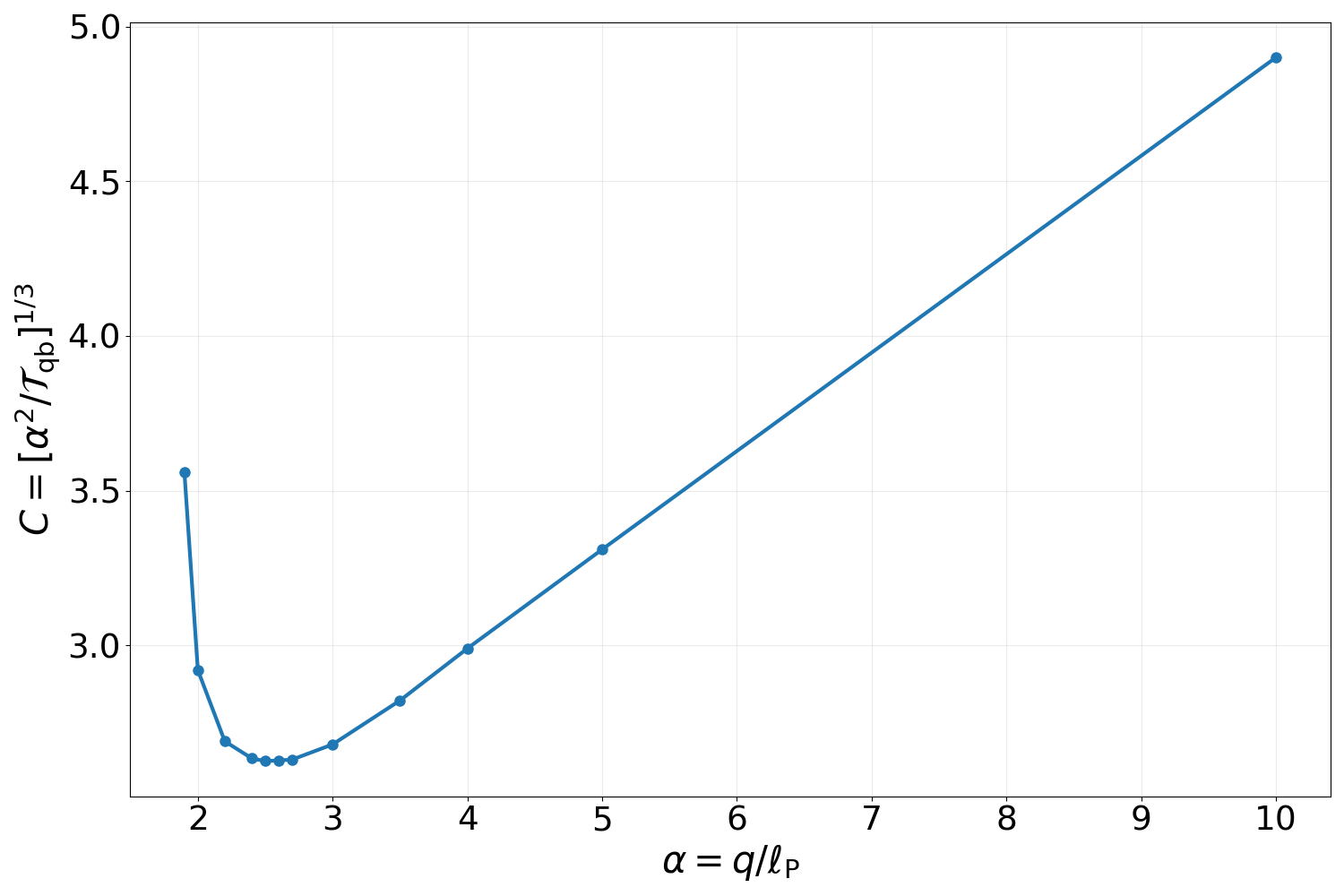}\\[-1mm]
{\small (b) Boundary-radius coefficient}
\end{minipage}
\caption{Graphical representation of Table~\ref{tab:schwarzschild_Tqb}.
(a) Critical dimensionless Schwarzschild tidal strength
$\mathcal{T}_{\rm qb}$ for loss of connected D-state support.
(b) The derived coefficient
$C=[\alpha^2/\mathcal{T}_{\rm qb}]^{1/3}$ in
$r_{\rm qb}=C(r_g\ell_{\rm P}^2)^{1/3}$.  The broad minimum in $C$ occurs
near $\alpha\simeq2.5$--2.7.}
\label{fig:schwarzschild_scan_plot}
\end{figure*}

Figure~\ref{fig:schwarzschild_scan_plot} shows the same scan graphically.
The upper panel displays the monotonic increase of the critical tidal
strength with $\alpha$, while the lower panel makes clear the shallow minimum
of the radius coefficient $C$ near $\alpha\simeq2.5$--2.7.

where the corresponding Schwarzschild quantum-boundary radius is
\begin{equation}
\begin{aligned}
 r_{\rm qb}(\alpha)
 &=\left[\frac{\alpha^2}{\mathcal{T}_{\rm qb}(\alpha)}\right]^{1/3}
  (r_g\ell_P^2)^{1/3}\\
 &=C(\alpha)(r_g\ell_P^2)^{1/3}.
\end{aligned}
\label{eq:schwarzschildrBQ}
\end{equation}
The smallest $r_{\rm qb}$ in the scan shown in
Table~\ref{tab:schwarzschild_Tqb} occurs for $\alpha\simeq2.5$,
for which $C\simeq2.63$, assuming $C_W$ is close to 1.  Spatial quanta close to the minimum allowed size
fail farther out because their intrinsic quantum fluctuations are already
near the D-state fragmentation threshold.  Very large quanta also fail
farther out because the tidal variation across a cell grows as $q^2$.
The competition produces an intermediate scale that penetrates deepest.

\subsection{Homogeneous Ricci-Induced Quantum Boundaries}
\label{sec:homogeneousBQ}

Spatially flat Friedmann geometry, $ds^2=-c^2dt^2+a^2(t)d\mathbf{x}^2$,
is conformally flat; for radiation, $C_{\alpha\beta\gamma\delta}=0$ and
$R=0$, so the finite-scale metric stress is purely trace-free Ricci.  In an
orthonormal frame,
$S_{\hat\mu\hat\nu}=-(\dot H/2c^2)\operatorname{diag}(3,1,1,1)$.
Radiation domination has $\dot H=-2H^2$, and hence
\begin{equation}
 S_{\hat\mu\hat\nu}
 =
 \frac{H^2}{c^2}\operatorname{diag}(3,1,1,1).
\label{eq:SFRWradiation}
\end{equation}
The dimensionless curvature across one spatial quantum is
$\mathcal{T}\equiv H^2q^2/c^2$.
Unlike the common factor $a^2(t)$, $\mathcal T$ changes the finite-scale
relation between timelike and spacelike geometry.

We tested this response with the gauge-reduced full four-metric ensemble:
four linearized de Donder conditions reduce the ten symmetric components
to six degrees of freedom, and a realization is Lorentzian only when the
full $4\times4$ metric has exactly one negative eigenvalue.  The local
finite-$q$ response model is
\begin{equation}
 s^{(R)}_{\hat\mu\hat\nu}
 =
 -C_Rq^2S_{\hat\mu\hat\nu}
 =
 -C_R\mathcal T\,\operatorname{diag}(3,1,1,1),
\label{eq:RicciResponse}
\end{equation}
The scans set $C_R=1$, but more generally determine the control variable
$\Theta\equiv C_R\mathcal T$.  A covariant derivation of $C_R$ from a
specified averaging or normal-coordinate prescription remains open, so the
absolute curvature and density scales below vary as $C_R^{-1}$.

For $\alpha=2.5$, the midpoint thresholds on $32^3$, $48^3$, and $64^3$
physical-$q$ lattices were $0.6839\pm0.0039$, $0.6865\pm0.0047$, and
$0.6843\pm0.0025$, giving the combined result
$\Theta_{\rm qb}=0.6846\pm0.0019$.  The $64^3$ scans gave
$\Theta_{\rm qb}=0.5955\pm0.0032$ for $\alpha=1.9$ and
$0.7983\pm0.0015$ for $\alpha=4$.  At extinction
$P_{\rm L}\simeq0.43$: finite differentiable islands
remain, but none winds around the periodic box.  The quoted uncertainties
measure numerical sampling of $\Theta$, not uncertainty in the physical
density, which is dominated by the unknown $C_R$.

From the Friedmann equation and
$\rho_{\rm P}=c^5/(\hbar G^2)$,
\begin{equation}
 \frac{\rho}{\rho_{\rm P}}
 =
 \frac{3\mathcal{T}}{8\pi\alpha^2}.
\label{eq:rhoFRWx}
\end{equation}
Thus, for $\alpha=2.5$,
\begin{equation}
 \boxed{\frac{\rho_{\rm qb}}{\rho_{\rm P}}
 \simeq\frac{1.31\times10^{-2}}{C_R}},
\label{eq:rhoBQhomogeneous}
\end{equation}
For comparison, the local Lorentzian fraction reaches $P_{\rm L}=0.5$ at
$\Theta\simeq0.65$, or $\mathcal T\simeq0.65/C_R$, corresponding to
$\rho/\rho_{\rm P}\simeq1.24\times10^{-2}/C_R$; periodic winding therefore
persists beyond the one-point 50-percent level.  The $\alpha=1.9$ and 4
boundary densities are respectively $1.97\times10^{-2}/C_R$ and
$5.96\times10^{-3}/C_R$.  These conditional values place homogeneous
signature failure near one percent of $\rho_{\rm P}$ only when $C_R$ is
of order unity.

This boundary is reached only if collapse enters the corresponding
curvature regime; pressure or rotation may delay, but cannot generically
prevent, collapse above the maximum stable mass.  Because
Eq.~(\ref{eq:SFRWradiation}) depends on $H^2$, the same local boundary
applies to contraction and the time-reversed expanding branch.  In the
latter interpretation it is a nucleation boundary, with no physical
Lorentzian manifold on its high-density side.

\section{Minimum Spatial Quanta in Higher Dimensions}

The preceding calculation was carried out in $3+1$ dimensions.  Since
candidate theories of quantum gravity frequently contain additional spatial
dimensions, it is useful to ask whether the signature-failure criterion
changes qualitatively in a spacetime of dimension $D=d+1$.  A full
$d$-dimensional lattice calculation would scale as $N^d$ and rapidly become
impractical.  We make a simpler one-point estimate of the
Lorentzian survival condition.

For $d$ spatial dimensions, the number of physical transverse-traceless
graviton polarizations is $N_{\rm pol}(d)=(d+1)(d-2)/2$.
Each polarization is quantized as an independent harmonic oscillator with
the usual canonical commutator.  With the $D$-dimensional Planck length
$\ell_D^{\,d-1}=\hbar G_D/c^3$ and physical spatial quantum
$q=\alpha\ell_D$, the canonical vacuum condition
$\Delta\pi=\omega\Delta v$ for a localized TT wavepacket of scale $q$,
with $\omega\sim c/q$, gives
$\Delta_g^2\propto\alpha^{-(d-1)}$.  Normalizing the localized packet so
that the $d=3$ case reproduces the four-dimensional result used above gives
\begin{equation}
 \Delta_g^2(\alpha,d)
 =
 \frac{4}{\pi}N_{\rm pol}(d)\,
 \frac{1}{\alpha^{d-1}}.
\label{eq:higherDdelta}
\end{equation}
For $d=3$ and $N_{\rm pol}=2$, Eq.~(\ref{eq:higherDdelta})
reduces to $\Delta_g^2=(8/\pi)\alpha^{-2}$.

At one point, the angularly averaged TT vacuum ensemble is an isotropic
Gaussian ensemble of traceless symmetric $d\times d$ matrices.  Let
$P_{\rm nonL}(\alpha,d)$ denote the probability that
$I+\chi$ is not positive definite.  A complete centered differentiable
neighborhood contains $3^d$ spatial quanta.  Neglecting correlations and
causal repair, a rough survival condition is therefore, $P_{\rm nonL}(\alpha,d)\sim 1/3^d$ at the onset of fragmentation. This is not the full percolation threshold; it is
a local statistical estimate intended to determine whether a
higher-dimensional solution is plausible without constructing an
$N^d$ lattice.

For representative spatial dimensions $d=3$, $5$, $9$, and $10$, this
one-point estimate gives $\alpha_{\min}\simeq1.77$, $1.88$, $1.57$, and
$1.51$, respectively.  The $d=3$ value is reasonably close to the full
correlated result $q_{\min}\simeq1.9\ell_{\rm P}$.  Because the present
higher-dimensional estimate omits the full correlated connectivity and
causal evolution, we expect these values to underestimate the true
$\alpha_{\min}$.  Conservatively, the minimum spatial quantum in each case
should therefore not lie below these values of $\alpha\ell_D$.

\section{Discussion}
\label{sec:discussion}

The calculations identify complementary curvature-driven routes to signature
failure: vacuum Weyl curvature produces anisotropic tidal stress, whereas
homogeneous radiation produces trace-free Ricci stress.  The Schwarzschild
case provides a possible microscopic endpoint before its classical tidal
deformation diverges.  The singularity theorems establish geodesic incompleteness under specified causal and energy assumptions, but they do not explain its physical origin or identify what lies at the incomplete endpoint \cite{Penrose1965,HawkingPenrose1970,Penrose1979,HawkingEllis1973}. Here the failure of continuation has a direct geometrical explanation. Geodesics end not because they encounter a physical point of infinite density, but because there is no spacetime beyond the quantum boundary ($\Psi_{qb}=0$) through which they could continue.

More generally, for fixed $q$, any evolution satisfying
$ (q/c)(\sigma_{\mu\nu}\sigma^{\mu\nu})^{1/2}\rightarrow\infty$
must cross the finite shear threshold for boundary formation.  The same
holds for unbounded tidal deformation on scale $q$.  Applications to
BKL-type evolution, black-hole mass inflation, and inhomogeneous
cosmological collapse are deferred to companion studies
\cite{shaya2026quantumboundaryblackhole,
shaya2026topologicalmultiverseselfconsistentextension}.

The Compton--Schwarzschild and connectivity scales are ordered.  Gravitational
closure becomes characteristic only below $\sqrt{2}\,\ell_{\rm P}$,
whereas periodic differentiable support is already lost at
$q_{\min}\simeq1.9\ell_{\rm P}$.  At the latter scale,
\begin{equation}
 \frac{r_s}{q}
 < 2\left(\frac{\ell_{\rm P}}{q_{\min}}\right)^2
 =\frac{2}{1.9^2}
 \simeq0.554 .
\label{eq:closure_ratio_qmin}
\end{equation}
Signature failure occurs first because the smallest eigenvalue of a
fluctuating symmetric metric can cross zero while each component remains
sub-unity; at threshold the rms of one diagonal component is about $0.31$.
Closure instead requires an order-unity localization-energy excursion.
Thus the manifold loses a differentiable Lorentzian state before a spatial quantum becomes
smaller than the region closed by its own characteristic energy.

\section*{Code Availability}

The numerical codes used in this work are supplied with the arXiv submission
and will is maintained in a public GitHub repository, https://github.com/eshaya/Signature.
\texttt{bq\_alpha\_stability\_cupy.py} implements the GPU TT-vacuum
generation, causal D-state evolution, and periodic-winding calculation used
to determine the minimum stable $\alpha$.  The same numerical machinery
underlies the Schwarzschild calculation after addition of the deterministic
electric-Weyl deformation
$\mathcal T\operatorname{diag}(2,-1,-1)$ to the spatial metric.
\texttt{homogeneous\_radiation\_qb\_Dstate.py} performs the gauge-reduced
full-four-metric scans for homogeneous trace-free Ricci curvature.

\begin{acknowledgments}
The author acknowledges the computational resources provided by the
University of Maryland. Generative artificial intelligence played a
significant role in developing the mathematical formulation and the
numerical code used to obtain the results. The conjectures here did not originate with the AI models.  The author independently
verified the scientific arguments, calculations, and conclusions.  In addition, various models were usefully engaged to judge the work, provide competing numerical codes, and suggested improvements.  The drafts were circulated periodically between Anthropic models Opus 5 and Fable 5.1, OpenAi 5.6 Sol  and Tera, and Gemini 3.1 Pro and 3.8 Flash.
\end{acknowledgments}

\bibliographystyle{apsrev4-2}
\bibliography{references_v4}

@article{Asgeirsson1937,
  author  = {{\'{A}}sgeirsson, Leifur},
  title   = {\"{U}ber eine {M}ittelwertseigenschaft von {L}\"{o}sungen homogener linearer partieller {D}ifferentialgleichungen 2.~{O}rdnung mit konstanten {K}oeffizienten},
  journal = {Mathematische Annalen},
  volume  = {113},
  pages   = {321--346},
  year    = {1937},
}

@article{CraigWeinstein2009,
  author  = {Craig, Walter and Weinstein, Steven},
  title   = {On Determinism and Well-Posedness in Multiple Time Dimensions},
  journal = {Proceedings of the Royal Society A},
  volume  = {465},
  pages   = {3023--3046},
  year    = {2009},
}

@article{DeWitt1967,
  author  = {DeWitt, Bryce S.},
  title   = {Quantum Theory of Gravity. {I}. {T}he Canonical Theory},
  journal = {Physical Review},
  volume  = {160},
  pages   = {1113--1148},
  year    = {1967},
  note    = {doi:10.1103/PhysRev.160.1113}
}

@article{John1938,
  author  = {John, Fritz},
  title   = {The Ultrahyperbolic Differential Equation with Four Independent Variables},
  journal = {Duke Mathematical Journal},
  volume  = {4},
  pages   = {300--322},
  year    = {1938},
}

@incollection{Penrose1979,
  author    = {Penrose, Roger},
  title     = {Singularities and Time-Asymmetry},
  booktitle = {General Relativity: An Einstein Centenary Survey},
  editor    = {Hawking, Stephen W. and Israel, Werner},
  publisher = {Cambridge University Press},
  address   = {Cambridge},
  year      = {1979},
  pages     = {581--638}
}

@incollection{Wheeler1968,
  author    = {Wheeler, John A.},
  title     = {Superspace and the Nature of Quantum Geometrodynamics},
  booktitle = {Battelle Rencontres: 1967 Lectures in Mathematics and Physics},
  editor    = {DeWitt, C\'ecile M. and Wheeler, John A.},
  publisher = {Benjamin},
  address   = {New York},
  year      = {1968},
  pages     = {242--307}
}

@article{ChoquetBruhat1952,
  author  = {Choquet-Bruhat, Yvonne},
  title   = {Th\'{e}or\`{e}me d'existence pour certains syst\`{e}mes d'\'{e}quations aux d\'{e}riv\'{e}es partielles non lin\'{e}aires},
  journal = {Acta Mathematica},
  volume  = {88},
  pages   = {141--225},
  year    = {1952}
}

@article{ChoquetBruhatGeroch1969,
  author  = {Choquet-Bruhat, Yvonne and Geroch, Robert},
  title   = {Global aspects of the {C}auchy problem in general relativity},
  journal = {Communications in Mathematical Physics},
  volume  = {14},
  number  = {4},
  pages   = {329--335},
  year    = {1969}
}

@article{Geroch1970,
  author  = {Geroch, Robert},
  title   = {Domain of dependence},
  journal = {Journal of Mathematical Physics},
  volume  = {11},
  number  = {2},
  pages   = {437--449},
  year    = {1970},
  note    = {doi:10.1063/1.1665157}
}

@book{HawkingEllis1973,
  author    = {Hawking, S. W. and Ellis, G. F. R.},
  title     = {The Large Scale Structure of Space-Time},
  publisher = {Cambridge University Press},
  address   = {Cambridge},
  year      = {1973}
}

@book{Wald1984,
  author    = {Wald, Robert M.},
  title     = {General Relativity},
  publisher = {University of Chicago Press},
  address   = {Chicago},
  year      = {1984}
}

@article{Dimock1980,
  author  = {Dimock, J.},
  title   = {Algebras of local observables on a manifold},
  journal = {Communications in Mathematical Physics},
  volume  = {77},
  pages   = {219--228},
  year    = {1980}
}

@article{KayWald1991,
  author  = {Kay, Bernard S. and Wald, Robert M.},
  title   = {Theorems on the uniqueness and thermal properties of stationary, nonsingular, quasifree states on spacetimes with a bifurcate {K}illing horizon},
  journal = {Physics Reports},
  volume  = {207},
  number  = {2},
  pages   = {49--136},
  year    = {1991}
}

@book{Wald1994,
  author    = {Wald, Robert M.},
  title     = {Quantum Field Theory in Curved Spacetime and Black Hole Thermodynamics},
  publisher = {University of Chicago Press},
  address   = {Chicago},
  year      = {1994}
}

@article{Grishchuk1974,
  author  = {Grishchuk, L. P.},
  title   = {Amplification of gravitational waves in an isotropic universe},
  journal = {Soviet Physics JETP},
  volume  = {40},
  pages   = {409},
  year    = {1975},
  note    = {[Zh. Eksp. Teor. Fiz. \textbf{67}, 825 (1974)]}
}

@article{MFB1992,
  author  = {Mukhanov, V. F. and Feldman, H. A. and Brandenberger, R. H.},
  title   = {Theory of cosmological perturbations},
  journal = {Physics Reports},
  volume  = {215},
  number  = {5--6},
  pages   = {203--333},
  year    = {1992}
}

@article{Wigner1958,
  author  = {Wigner, Eugene P.},
  title   = {On the distribution of the roots of certain symmetric matrices},
  journal = {Annals of Mathematics},
  volume  = {67},
  number  = {2},
  pages   = {325--327},
  year    = {1958}
}

@article{Dyson1962,
  author  = {Dyson, Freeman J.},
  title   = {Statistical theory of the energy levels of complex systems. {I}},
  journal = {Journal of Mathematical Physics},
  volume  = {3},
  number  = {1},
  pages   = {140--156},
  year    = {1962}
}

@book{Mehta2004,
  author    = {Mehta, Madan Lal},
  title     = {Random Matrices},
  edition   = {3rd},
  publisher = {Elsevier/Academic Press},
  address   = {Amsterdam},
  series    = {Pure and Applied Mathematics},
  volume    = {142},
  year      = {2004}
}

@article{Mead1964,
  author  = {Mead, C. A.},
  title   = {Possible Connection Between Gravitation and Fundamental Length},
  journal = {Physical Review},
  volume  = {135},
  pages   = {B849--B862},
  year    = {1964},
  doi     = {10.1103/PhysRev.135.B849}
}

@article{Garay1995,
  author  = {Garay, Luis J.},
  title   = {Quantum Gravity and Minimum Length},
  journal = {International Journal of Modern Physics A},
  volume  = {10},
  number  = {2},
  pages   = {145--165},
  year    = {1995},
  doi     = {10.1142/S0217751X95000085},
  eprint  = {gr-qc/9403008}
}

@article{Hossenfelder2013,
  author  = {Hossenfelder, Sabine},
  title   = {Minimal Length Scale Scenarios for Quantum Gravity},
  journal = {Living Reviews in Relativity},
  volume  = {16},
  pages   = {2},
  year    = {2013},
  doi     = {10.12942/lrr-2013-2},
  eprint  = {1203.6191},
  archivePrefix = {arXiv},
  primaryClass  = {gr-qc}
}

@article{GrossMende1988,
  author  = {Gross, David J. and Mende, Paul F.},
  title   = {String Theory Beyond the Planck Scale},
  journal = {Nuclear Physics B},
  volume  = {303},
  number  = {3},
  pages   = {407--454},
  year    = {1988},
  doi     = {10.1016/0550-3213(88)90390-2}
}

@article{Bombelli1987,
  author  = {Bombelli, Luca and Lee, Joohan and Meyer, David and Sorkin, Rafael D.},
  title   = {Space-Time as a Causal Set},
  journal = {Physical Review Letters},
  volume  = {59},
  number  = {5},
  pages   = {521--524},
  year    = {1987},
  doi     = {10.1103/PhysRevLett.59.521}
}

@article{RovelliSmolin1995,
  author  = {Rovelli, Carlo and Smolin, Lee},
  title   = {Discreteness of Area and Volume in Quantum Gravity},
  journal = {Nuclear Physics B},
  volume  = {442},
  number  = {3},
  pages   = {593--619},
  year    = {1995},
  doi     = {10.1016/0550-3213(95)00150-Q}
}

@article{Ambjorn2005,
  author  = {Ambjorn, Jan and Jurkiewicz, Jerzy and Loll, Renate},
  title   = {Reconstructing the Universe},
  journal = {Physical Review D},
  volume  = {72},
  pages   = {064014},
  year    = {2005},
  doi     = {10.1103/PhysRevD.72.064014}
}

@misc{shaya2026topologicalmultiverseselfconsistentextension,
      title={The Topological Multiverse as the Self-Consistent Extension of Everettian QM to Quantum Gravity},
      author={Edward J Shaya},
      year={2026},
      eprint={2512.03251},
      archivePrefix={arXiv},
      primaryClass={gr-qc},
      url={https://arxiv.org/abs/2512.03251},
}

@misc{shaya2026quantumboundaryblackhole,
      title={The Quantum Boundary of Black Hole Interiors: Termination of the Sum over Geometries at Planck Curvature},
      author={Edward J. Shaya},
      year={2026},
      eprint={2606.15423},
      archivePrefix={arXiv},
      primaryClass={gr-qc},
      url={https://arxiv.org/abs/2606.15423},
}

@article{Penrose1965,
  author  = {Penrose, Roger},
  title   = {Gravitational Collapse and Space-Time Singularities},
  journal = {Physical Review Letters},
  volume  = {14},
  pages   = {57--59},
  year    = {1965},
  note    = {doi:10.1103/PhysRevLett.14.57}
}

@article{HawkingPenrose1970,
  author  = {Hawking, S. W. and Penrose, R.},
  title   = {The Singularities of Gravitational Collapse and Cosmology},
  journal = {Proceedings of the Royal Society of London A},
  volume  = {314},
  pages   = {529--548},
  year    = {1970},
  note    = {doi:10.1098/rspa.1970.0021}
}

@article{HartleHawking1983,
  author  = {Hartle, James B. and Hawking, S. W.},
  title   = {Wave Function of the Universe},
  journal = {Physical Review D},
  volume  = {28},
  pages   = {2960--2975},
  year    = {1983},
  doi     = {10.1103/PhysRevD.28.2960}
}

@article{EllisSumerukCouleHellaby1992,
  author  = {Ellis, George F. R. and Sumeruk, Ariel and Coule, D. H. and Hellaby, Charles},
  title   = {Change of Signature in Classical Relativity},
  journal = {Classical and Quantum Gravity},
  volume  = {9},
  number  = {6},
  pages   = {1535--1554},
  year    = {1992},
  doi     = {10.1088/0264-9381/9/6/011}
}

@article{DrayManogueTucker1991,
  author  = {Dray, Tevian and Manogue, Corinne A. and Tucker, Robin W.},
  title   = {Particle Production from Signature Change},
  journal = {General Relativity and Gravitation},
  volume  = {23},
  number  = {8},
  pages   = {967--971},
  year    = {1991},
  doi     = {10.1007/BF00756915}
}

@article{Hayward1992,
  author  = {Hayward, Sean A.},
  title   = {Signature Change in General Relativity},
  journal = {Classical and Quantum Gravity},
  volume  = {9},
  pages   = {1851--1862},
  year    = {1992},
  doi     = {10.1088/0264-9381/9/8/007}
}

@article{KossowskiKriele1993,
  author  = {Kossowski, Marek and Kriele, Martin},
  title   = {Signature Type Change and Absolute Time in General Relativity},
  journal = {Classical and Quantum Gravity},
  volume  = {10},
  pages   = {1157--1164},
  year    = {1993}
}

\end{document}